\newcommand{\eqn}[1]{(\ref{#1})}
\newcommand{\ft}[2]{{\textstyle\frac{#1}{#2}}}
\def\be{\begin{equation}}
\def\ee{\end{equation}}
\def\bea{\begin{eqnarray}}
\def\eea{\end{eqnarray}}
\def\nn{\nonumber}

\def\nn{{\nonumber}}

\renewcommand{\a}{\alpha}
\renewcommand{\b}{\beta}
\def\vep{\varepsilon}                         

\renewcommand{\d}{\delta}
\newcommand{\pa}{\partial}
\newcommand{\g}{\gamma}
\newcommand{\G}{\Gamma}
\newcommand{\e}{\epsilon}

\renewcommand{\l}{\lambda}
\newcommand{\m}{\mu}
\newcommand{\n}{\nu}

\newcommand{\s}{\sigma}

\renewcommand{\o}{\omega}

\def\IZ{{\hbox{{\rm Z}\kern-.4em\hbox{\rm Z}}}}

\def\bigone{{\hbox{1\kern -.23em{\rm l}}}}
\newcommand{\NPB}[3]{{Nucl.\ Phys.} {\bf B#1} (#2) #3}

\newcommand{\PRD}[3]{{ Phys.\ Rev.}\ {\bf D#1} (#2) #3}
\newcommand{\PLB}[3]{{Phys.\ Lett.}\ {\bf B#1} (#2) #3}
\newcommand{\JHEP}[3]{{JHEP} {\bf #1} (#2) #3}

\documentclass{article}

\begin{document}
\begin{flushright}
THU-99/06\\[1mm]
{\tt hep-th/9902149}
\end{flushright}

\begin{center}
{\large {\bf SUPERMEMBRANES IN CURVED SUPERSPACE \\
AND NEAR-HORIZON GEOMETRIES}}
\vskip 8mm
B. DE WIT
\vskip 3mm
Institute for Theoretical Physics, Utrecht University\\
Princetonplein 5, 3508 TA Utrecht, The Netherlands\\
E-mail: bdewit@phys.uu.nl
\vskip 3mm
{\bf Abstract}
\end{center}
\noindent
We give a description of supermembranes in nontrivial target-space
geometries. A special class are the $AdS_4\times S^7$ and 
$AdS_7\times S^4$ spaces that have the maximal number of 32
supersymmetries. 
\\   
Lecture presented at the 22-nd Johns Hopkins Workshop, {\it Novelties
in String Theory}, G\"oteborg, August
20-22, 1998. 

\section{Introduction}
The original study of supermembranes  \cite{BST} was motivated by the desire to
find a consistent quantum-mechanical extension of 11-dimensional
supergravity  \cite{CJS} along the same lines that string theory defines
a quantum-mechanical extension of 10-dimensional supergravity theories. 
In 11 spacetime dimensions the supermembrane can consistently
couple to a superspace background that satisfies a number of
constraints which are equivalent to the supergravity equations
of motion. The supermembrane action
can also exist in $4,5$ and 7 spacetime dimensions, 
just as the Green-Schwarz superstring  \cite{GS84} is
classically consistent in $3,4,6$ and 10 
dimensions.  Guided by string theory it was natural to 
expect that the massless states of the supermembrane would 
correspond to those of 11-dimensional 
supergravity. However, the supermembrane mass spectrum turned out to
be continuous  \cite{DWLN} and in that situation it was cumbersome to
directly prove or disprove the  possible existence of massless
states  \cite{DWHN,DWN}. The
difficulties in making sense of the supermembrane mass spectrum  
and the fact that 11-dimensional supergravity seemed to have no place
in string theory, formed an obstacle for subsequent development of the
theory. More recently, however, interest in supermembranes was 
rekindled by the realization that 11-dimen\-sion\-al supergravity 
does have its role to play as the long-distance approximation to
M-theory  \cite{HT,Townsend,witten3,horvw}.  M-theory is the 
conjectured framework for unifying  
all five superstring theories and 11-dimensional supergravity. It
turns out that supermembranes, M-theory and super matrix theory are all
intricately related. 

An important observation was that it is possible to regularize 
the supermembrane in terms of a super matrix model based on some 
finite group, such as U($N$). In the limit of infinite $N$ one 
then recovers the supermembrane  \cite{DWHN}. These supersymmetric
matrix models were  
constructed long ago  \cite{CH} and can be obtained from 
supersymmetric Yang-Mills theories in the zero-volume 
limit. More recently it was realized that these models describe 
the short-distance behaviour of $N$ Dirichlet particles  \cite{boundst}. The 
continuity of the spectrum is then understood directly in terms of 
the spectrum of $N$-particle states. A bold conjecture was that the 
super matrix models capture the degrees of freedom of
M-theory  \cite{BFSS}. In the large-$N$ limit, where one considers the  
states with an infinite number of particles, the supermembranes 
should then re-emerge. Furthermore 
there is new evidence meanwhile that the supermembrane has massless
states, which will presumably correspond to the states of 11-dimensional
supergravity, although proper asymptotic states do not exist. The
evidence is based on the matrix model regularization of the
supermembrane  \cite{mboundstates}. For fixed value of $N$ the 
existence of such states was foreseen on the basis of identifying
the Kaluza-Klein states of M-theory compactified on $S^1$ with the
Dirichlet particles and their bound states in type-IIA string
theory. 

{From} this viewpoint it is natural to consider the supermembrane in
curved backgrounds associated with 11-dimensional supergravity and
it is the purpose of this talk to report on progress in this
direction. Such backgrounds 
consist of a nontrivial metric, a three-index gauge
field and a gravitino field. This provides us with an
action that transforms as a scalar under the combined (local) 
supersymmetry transformations of the background fields and the
supermembrane embedding coordinates. Here it is important to
realize that the supersymmetry transformations of the embedding
coordinates will themselves depend on the background. When
the background is supersymmetric, then the action will be
supersymmetric as well. 
In the light-cone formulation this action leads to models
invariant under area-preserving diffeomorphisms, which in
certain situations can be approximated by matrix models 
in curved backgrounds. The area-preserving diffeomorphisms
are then replaced by a finite group, such as U$(N)$, but 
target-space diffeomorphisms are no longer manifestly
realized. Matrix models in curved space have already been the subject 
of independent studies  \cite{oldBG}. Also in view of the relation
between near-horizon  
geometries and conformal field theories  \cite{maldacena} 
interesting classes of backgrounds are the ones where the target 
space factorizes locally into the product of an anti-de Sitter space
and some compact space. Examples of these are the $AdS_4\times S^7$ and
$AdS_7\times S^4$ backgrounds that we discuss at the end of the talk.

\section{Supermembranes}                         
Fundamental supermembranes can be described in terms of actions 
of the Green-Schwarz type, possibly in a nontrivial but 
restricted (super)spacetime background  \cite{BST}. Such actions exist 
for supersymmetric $p$-branes, where $p= 0, 1,\ldots, d-1$ defines the
spatial dimension of the brane. Thus for
$p=0$ we have a superparticle, for $p=1$ a superstring, for
$p=2$ a supermembrane, and so on. The
dimension of spacetime in which the superbrane can live is
very restricted. These restrictions arise from the 
fact that the action contains a 
Wess-Zumino-Witten term, whose supersymmetry depends sensitively
on the spacetime dimension. If the coefficient of this term
takes a particular value then the action possesses an additional
fermionic gauge symmetry, the so-called $\kappa$-symmetry.
This symmetry is necessary to
ensure the matching of (physical) bosonic and fermionic
degrees of freedom.
In the following we restrict ourselves to 
supermembranes (i.e., $p=2$) in 11 dimensions.  

The 11-dimensional supermembrane is written in terms of 
superspace embedding coordinates 
$Z^M(\zeta)=(X^\mu(\zeta),\theta^\alpha(\zeta))$, which are functions of 
the three world-volume coordinates $\zeta^i$ ($i = 0,1,2$). It 
couples to the superspace geometry of 11-dimensional supergravity,
encoded by the supervielbein $E_M{}^{\!A}$ and the antisymmetric tensor
gauge superfield $B_{MNP}$, through the action\footnote{%
  Our notation and conventions are as follows. Tangent-space indices are
  $A=(r,a)$, whereas curved indices are denoted by
  $M=(\mu,\alpha)$. Here $r,\mu$ refer to  commuting and
  $a,\alpha$ to anticommuting coordinates. Moreover we take 
  $\epsilon_{012}=-\epsilon^{012}=1$. } %
\be
S[Z(\zeta)] = \int {\rm d}^3\zeta \;\Big[- \sqrt{-g(Z(\zeta))} - \ft16
\varepsilon^{ijk}\, \Pi_i^A \Pi_j^B\Pi_k^C \,B_{CBA}(Z(\zeta)) 
\,\Big]\,,
\label{supermem}
\ee
where $\Pi^A_i = \pa Z^M/\pa\zeta^i \; E_M{}^{\!A}$ is the pull-back of
the supervielbein to the membrane worldvolume. Here the 
induced metric equals $g_{ij} =\Pi^r_i \Pi^s_j \,\eta_{rs}$, with
$\eta_{rs}$ being the constant Lorentz-invariant metric. This action
is invariant under local fermionic $\kappa$-transformations 
alluded to above,
given that certain constraints on the background fields hold, which are 
equivalent to the  equations of motion of 11-dimensional 
supergravity  \cite{BST}.

Flat superspace is characterized by
\be\label{flatssquantities}
\begin{array}{rclcrcl}
E_\mu{}^r &=& \d_\mu{}^r \, , & & E_\mu{}^a &=&0  \, , \\
E_\alpha{}^a &=& \delta_\alpha{}^a  \, ,& & 
E_\alpha{}^r &=& -(\bar\theta \Gamma^r)_{\alpha} \, , \\
B_{\mu\n\alpha} &=& (\bar\theta\Gamma_{\m\n})_{\alpha}\,
, &&
B_{\m\a\b} &=& 
(\bar\theta\Gamma_{\m\n})_{(\a}\,(\bar\theta\Gamma^\n)_{\b)} \,, \\
 B_{\a\b\g} &=& (\bar\theta\Gamma_{\m\n})_{(\a}\, 
(\bar\theta\Gamma^\m)_{\b}\, 
(\bar\theta\Gamma^\n)_{\g)}\,,&\hspace{6.2mm}  &
B_{\mu\nu\rho} &=&0  \, .  
\end{array}
\ee
The gamma matrices 
are denoted by $\Gamma^r$; gamma matrices 
with more than one index denote antisymmetrized products of gamma
matrices with unit weight. In flat  
superspace the supermembrane Lagrangian, written in 
components, reads (in the notation and conventions of   \cite{DWHN}),  
\bea
{\cal L} &=&- \,\sqrt{-g(X,\theta)}\nonumber\\ 
&&- \varepsilon^{ijk} \,
\bar\theta\Gamma_{\mu\nu}\partial_k\theta \Big [{\textstyle{1\over 2}}
\,\partial_i X^\mu (\partial_j X^\nu +
\bar\theta\Gamma^\nu\partial_j \theta) + {\textstyle{1\over 6}}
\,\bar\theta\Gamma^\mu\partial_i\theta\;
\bar\theta\Gamma^\nu\partial_j\theta \Big] \,, \qquad{}
\label{action}
\eea
The target space can have compact dimensions which permit winding
membrane states. In flat superspace the induced metric,
\be
g_{ij} = (\partial_iX^\mu + \bar\theta\Gamma^\mu
\partial_i\theta) (\partial_jX^\nu +
\bar\theta\Gamma^\nu
\partial_j\theta) \,\eta_{\mu\nu}\,,
\ee
is supersymmetric. Therefore the first term in 
\eqn{action} is trivially invariant under spacetime
supersymmetry,
\be
\d X^\mu = -\bar\e \G^\m\theta \,, \qquad \d\theta = \e\,.
\ee 
In $4$, $5$, $7$, or $11$ spacetime dimensions the 
second term in the action proportional to $\varepsilon^{ijk}$ is also 
supersymmetric (up to a total divergence) and the full action is 
invariant under  $\kappa$-symmetry.
 
Let us now consider the supermembrane action for nontrivial
backgounds, such as  
those induced by a nontrivial target-space metric, a target-space 
tensor field and a target-space gravitino field, corresponding to 
the fields of (on-shell) 11-dimensional supergravity. This
background can in principle be cast into superspace form by a
procedure known as `gauge completion'. For 11-dimensional 
supergravity, the first steps of this procedure were carried 
out long ago  \cite{CF} and recently  \cite{backgr} the results were
determined to second order in the fermionic coordinates $\theta$. 

To elucidate the generic effects of nontrivial backgrounds for
membrane theories, let us confine ourselves for the moment 
to the purely bosonic theory and present
the light-cone formulation of the membrane in a 
background consisting of the metric $G_{\mu\nu}$ and the tensor 
gauge field $C_{\mu\nu\rho}$. In the subsequent sections we will 
include the fermionic coordinates.
The Lagrangian density for the bosonic membrane follows directly 
{}from \eqn{supermem},  
\begin{equation}
{\cal L} = -\sqrt{-g} - \ft{1}{6}\varepsilon^{ijk} \partial_i X^\mu\,
\partial_j
X^\nu \,\partial_k X^\rho\, C_{\rho\nu\mu} \, ,
\end{equation}
where $g_{ij}= \pa_i X^\m \,\pa_j X^\n \,\eta_{\m\n}$.  
In the light-cone formulation, the coordinates are decomposed in 
the usual fashion as $(X^+,X^-,X^a)$ with  
$a=1\ldots 9$. Furthermore we use the diffeomorphisms in the 
target space to bring the metric in a convenient form  \cite{GS},
\begin{equation}
\label{metricgauge}
G_{--}=G_{a-}=0 \, .
\end{equation}
Just as for a flat target space, we identify the time coordinate
of the target space with the world-volume time, by imposing the
condition  $X^+ = \tau$. Moreover we denote the spacesheet
coordinates of the membrane by $\sigma^r$, $r=1,2$.
Following the same steps as for the membrane in flat 
space, one arrives at a Hamiltonian formulation of 
the theory in terms of coordinates and momenta. These
phase-space variables are subject to a constraint, which 
takes the same form as for the membrane theory in flat space, namely,  
\begin{equation}
\phi_r = P_a \,\partial_r X^a + P_- \,\partial_r X^- \approx 0\, .
\label{constraintmem}
\end{equation}
Of course, the definition of the momenta in terms of the 
coordinates and their derivatives does involve the 
background fields, but at the end all explicit dependence 
on the background cancels out in the phase-space constraints. 

The gauge choice $X^+=\tau$ still allows for $\tau$-dependent 
reparametrizations of the world-space coordinates
$\sigma^r$
In addition there remains the freedom of performing tensor gauge
transformations of the target-space three-form $C_{\mu\nu\rho}$.
In order to write the theory as a gauge theory of
area-preserving diffeomorphisms it is desirable to obtain a Hamiltonian
which is polynomial in momenta and coordinates. Obviously one wants
the background fields to be independent of $X^\pm$. With suitable
choices for the gauge condition  one thus derives  \cite{backgr},
\bea
H&\!=\!& \int {\rm d}^2\sigma\, \bigg \{
\frac{G_{+-}}{P_- - C_-}\bigg[\ft{1}{2}\Big(P_a-C_a-\frac{P_--C_-}{G_{+-}}
\, G_{a+}\Big)^2+ \ft{1}{4}
(\varepsilon^{rs}\, \partial_r X^a\, \partial_s X^b )^2\bigg]\nonumber\\
&&\hspace{15mm}  -\frac{P_--C_-}{2\, G_{+-}}\, G_{++}- C_+\nn\\
&&\hspace{15mm} +{1\over{P_--C_-}} \Big[ \varepsilon^{rs} \pa_r
X^a\pa_s X^b \,P_a\,C_{+-b} + C_-\,C_{+-}\Big]  \bigg \} \,, \label{BGham}
\eea
where $P_--C_-$ equals a constant times $\sqrt w$ and $C_{-ab}$ has
been set to a constant by a choice of gauge. Furthermore we made use
of the definitions,   
\bea
C_a &=& - \varepsilon^{rs} \partial_r X^- \partial_s X^b \,C_{-ab} +
\ft{1}{2}
\varepsilon^{rs}\partial_r X^b \partial_s X^c \,C_{abc} \, ,      
\nonumber \\
C_{\pm} &=& \ft{1}{2}\varepsilon^{rs}\partial_r X^a \partial_s X^b
\,C_{\pm ab}\,, \qquad 
C_{+-} = \varepsilon^{rs}\partial_r X^- \partial_s X^a\,
C_{+-a}\,.
\eea

At this point one can impose further gauge choices and set
$G_{+-}=1$ and $C_{+-a}=0$. Taking also $C_{-ab}=0$ the
corresponding Hamiltonian can be cast in Lagrangian form 
in terms of a gauge theory of area-preserving
diffeomorphisms  \cite{DWPP3}, 
\bea
{w}^{-1/2}\, {\cal L} &=& \ft{1}{2} (D_0 X^a)^2 + D_0 X^a \left(
\ft{1}{2} C_{abc} \{ X^b, X^c \} + G_{a+} \right) \nonumber \\
&&- \ft{1}{4}\{ X^a, X^b \}^2 + \ft12{G_{++}} + \ft{1}{2}
C_{+ab} \{ X^a, X^b \} \, , \label{BGlagr}
\eea
where the derivative is covariant with respect to area-preserving
diffeomorphisms. For convenience we have set $(P_-)_0 =1$.  
In the case of compact dimensions, it may not always be possible to 
set $C_{+-a}$ and $C_{-ab}$ to zero, although they can be 
restricted to constants. One can still follow the same procedure as 
above, but the
Lagrangian then depends explicitly on $X^-$, a feature that was
already exhibited earlier for the winding membrane. However, in the
case at hand, the constraint makes the resulting expression for $X^-$
extremely nontrivial. The 
antisymmetric constant matrix $C_{-ab}$ was conjectured to 
play a role for the matrix model compactification on a
noncommutative torus  \cite{newBG}. It should be interesting to
see what the role is of \eqn{BGham} in this context. 

With a reformulation of the membrane in background  fields as a 
gauge theory of area-preserving diffeomorphisms at one's
disposal, one may consider its regularization through a matrix
model by truncating 
the mode expansion for coordinates and momenta. This leads to a
replacement of Poisson 
brackets by commutators, integrals by traces and products of
commuting fields by symmetrized products of the corresponding 
matrices. At that point the original target-space covariance
is affected, as the matrix reparametrizations in terms of
symmetrized products of matrices do not possess a consistent
multiplication structure; 
this is just one of the underlying difficulties in the
construction of matrix models in curved space  \cite{oldBG}. 
With the above results at hand one may study interactions between
membranes by considering the behaviour of a test membrane in a
background field induced by another membrane  \cite{interactions}.  

\section{Superspace backgronds}
It is possible to express superspace backgrounds in terms of the
component fields of 11-dimensional supergravity, by means of a
technique called gauge completion. We refer to ref. 18 
for the details. The superspace geometry with coordinates
$Z^M=(x^\m,\theta^\a)$ is encoded in the supervielbein 
$E_M{}^{\!A}$ and a spin-connection field $\Omega_M{}^{\!AB}$. In
what follows we will not pay much 
attention to the spin-connection, which is not an independent
field. Furthermore we have an antisymmetric 
tensor gauge field $B_{MNP}$, subject to tensor gauge transformations, 
\be
\d B_{MNP} = 3\,\partial_{[M} \Xi_{NP]}\, .
\ee 
Unless stated otherwise the derivatives with respect to $\theta$
are always left derivatives. 

Under superspace diffeomorphisms corresponding to $Z^M\to
Z^M+\Xi^M(Z)$, the super-vielbein and tensor gauge field
transform as  
\bea
\delta E_M{}^{\!A} &=& \Xi^N \partial_N E_M{}^{\!A} + \partial_M \Xi^N
E_N{}^{\!A} \,,\nonumber \\ 
\delta B_{MNP} &=& \Xi^Q \partial_Q B_{MNP} 
+ 3\,\partial_{[M} \Xi^Q B_{\vert Q\vert NP]}\,.
\eea
Tangent-frame rotations are $Z$-dependent Lorentz
transformations that act on the vielbein according to
\be
\delta E_M{}^A =  \ft{1}{2}
(\Lambda^{rs}L_{rs})^A{}_{\!B}\, E_M{}^{\!B}\,,
\ee
where the Lorentz generators $L_{rs}$ are defined by
\be
\ft{1}{2}(\Lambda^{rs} L_{rs})^t{}_u = \Lambda^t{}_u \, ,\qquad
\ft{1}{2}(\Lambda^{rs} L_{rs})^a{}_b = \ft{1}{4}\Lambda^{rs}
(\Gamma_{rs})^a{}_b\, . 
\ee

The superspace that we are dealing with is not unrestricted but 
is subject to certain constraints and gauge
conditions. Furthermore, we will not describe an
off-shell situation 
as all superfields will be expressed entirely in terms of 
the three component fields of on-shell 11-dimensional supergravity,
the elfbein $e_\m{}^r$, the antisymmetric tensor gauge field
$C_{\m\n\rho}$ and the gravitino field $\psi_\m$. As a result of
these restrictions the residual symmetry transformations are confined to 
11-dimensional diffeomorphisms with parameters $\xi^\m(x)$, 
local Lorentz transformations with parameters $\l^{rs}(x)$, 
tensor-gauge transformations with parameters $\xi_{\m\n}(x)$ and
local supersymmetry transformations with parameters $\e(x)$. 
To derive how the superfields are
parametrized in terms of the component fields it is
necessary to also determine the form of the superspace
transformation parameters, $\Xi^M$,
$\Lambda^{rs}$ and $\Xi_{MN}$, that generate the supersymmetry
transformations. Here it is important to realize
that we are dealing with a gauge-fixed situation. For that reason
the superspace parameters depend on both the $x$-dependent
component parameters defined above as well as on the component 
fields. 
This has the consequence that local
supersymmetry transformations reside in the superspace
diffeomorphisms, the Lorentz transformations and the
tensor gauge transformations. Thus, when considering 
supersymmetry variations of the various fields, one must in
principle include each of the three possible superspace
transformations. 

For further details we refer to ref.18 
and we proceed directly to the results. 
For the supervielbein $E_M{}^{\!A}$ the following expressions were
found,  
\bea
E_\m{}^{\!r} &=& e_\m{}^r + 2\, \bar\theta\,\G^r \psi_\m
\nonumber \\
&& 
+ \bar\theta\,\G^r\Big[ -\ft14
\,\hat\omega_\m{}^{\!st}\G_{st} + T_\m{}^{\!\n\rho\s\lambda}\,\hat 
F_{\n\rho\s\lambda}\Big]\theta 
+  {\cal O}(\theta^3)\,, \nonumber \\
E_\m{}^{\!a} &=& \psi_\m{}^{\!a} - \ft14\,
\hat\o_\m^{rs}\,(\G_{rs} \theta)^a + 
(T_\m{}^{\!\n\rho\s\lambda}\theta)^a\, \hat F_{\n\rho\s\lambda} +
{\cal O}(\theta^2)\,, \nonumber \\
E_\a{}^{\!r} &=& -(\bar\theta\,\G^r)_\a +  {\cal O}(\theta^3)\,,
\nonumber \\  
E_\a{}^{\!a} &=& \d^a_\a + M_\a{}^{\!a} + {\cal O}(\theta^3)\,,
\label{svielbein}
\eea
where $M_\a{}^{\!a}$ characterizes the $\hat
F\theta^2$-contributions, which have not been evaluated. Observe that
$E_\m{}^{\!a}$ was determined only up to 
terms of order $\theta^2$. 
The result for the tensor field $B_{MNP}$ reads as follows,
\bea
B_{\m\n\rho} &=& C_{\m\n\rho} -6\, \bar \theta
\G_{[\m\n}\psi_{\rho]}
-3\,\bar\theta\,\G_{[\m\n}\Big[-\ft14\hat\omega_{\rho]}{}^{\!rs}\,\G_{rs} 
+ T_{\rho]}{}^{\!\s\lambda\kappa\tau}\,\hat
F_{\s\lambda\kappa\tau}\Big]\theta  \nonumber\\
&&  - 12\,
\bar\theta\,\G_{\s[\m} \psi_{\n}\;\bar\theta \,\G^\s \psi_{\rho]}
  +  {\cal O}(\theta^3) \,,\nonumber\\
B_{\m\n\a} &=& (\bar \theta \,\G_{\m\n})_\a - \ft83 \bar\theta\,\G^\rho
\psi_{[\m}\;(\bar\theta\,\G_{\n]\rho})_\a +\ft43
(\bar\theta\,\G^\rho)_\a \;\bar\theta\,\G_{\rho[\m}\psi_{\n]}
+  {\cal O}(\theta^3) \,, \nonumber\\ 
B_{\m\a\b} &=& (\bar\theta\,\G_{\m\n})_{(\a}\,
(\bar\theta\,\G^\n)_{\b)} 
+  {\cal
O}(\theta^3) \, ,  \nonumber \\
B_{\a\b\gamma} &=& (\bar\theta\Gamma_{\m\n})_{(\a}\, 
(\bar\theta\Gamma^\m)_{\b}\, 
(\bar\theta\Gamma^\n)_{\g)}
+  {\cal O}(\theta^3) \,. 
\label{stensor}
\eea
For completeness we included the $\theta^3$-term in
$B_{\a\b\gamma}$ which were already known from the flat-superspace
results \eqn{flatssquantities}. 

Then we turn to some of the transformation parameters. 
The supersymmetry transformations consistent with the fields
specified above, are generated by superspace diffeomorphisms,
local Lorentz transformations and tensor gauge
transformations. We only present the parameters for the 
superspace diffeomorphisms here and refer to the literature for more
complete results  \cite{backgr},
\bea
\Xi^\m(\e) &=& \bar\theta\, \G^\m \e -\bar \theta\,
\G^\n\e\;\bar\theta\,\G^\m\psi_\n
+ {\cal O}(\theta^3)\,, \nonumber\\ 
\Xi^\a(\e) &=& \e^\a -\bar\theta \,\G^\m\e\,\psi_\m^\a \nonumber \\
&&+ \bar \theta\,\G^\n\e\; \bar \theta\,
\G^\mu\psi_\n\,\psi_\m{}^{\!\a} +\ft14   \bar\theta\,\G^\n\e \,
\hat\omega^{rs}_\n\,(\G_{rs}\theta)^\a  + \e^\b \,N_\b{}^{\!\a} 
+  {\cal O}(\theta^3)\,, \label{sdiffs}
\eea 
where $N_\b{}^{\!\a}$ encodes unknown terms proportional to $\hat F
\theta^2$. 

Substituting the above results into the initial supermembrane action
\eqn{supermem} we obtain the action for a supermembrane in a
supergravity background to order $\theta^2$. While the original action
was manifestly covariant 
under independent superspace diffeomorphisms, tangent-space
Lorentz transformations  and tensor gauge
transformations, this is no longer the case and one has to
restrict oneself to the superspace 
transformations corresponding to the component supersymmetry,
general-coordinate, local Lorentz and tensor gauge 
transformations. When writing \eqn{supermem} in components,
utilizing the expressions found above, one
thus obtains an action that is covariant under the restricted
superspace 
diffeomorphisms \eqn{sdiffs} acting on the superspace
coordinates $Z^M=(X^\m,\theta^\a)$ (including the spacetime
arguments of the background fields) combined with usual
transformations on the component fields. Note that the result 
does not constitute an invariance. Rather it implies that 
actions corresponding to two different 
sets of background fields that are equivalent by a component
gauge transformation, are the same modulo a reparametrization of
the supermembrane embedding coordinates. In order to be precise let us
briefly turn to an example and consider the action of a particle
moving in a curved spacetime background with metric $g_{\m\n}$, 
\be 
S[X^\m, g_{\m\n}(X)]  = - m \int {\rm d}t\; \sqrt{- g_{\m\n}(X(t))\,\dot
X^\m(t)\,\dot X^\n(t)}\,. \label{particle}
\ee
This action, which is obviously invariant under world-line
diffeomorphisms, satisfies $S[X^{\prime\,\m}, 
g^\prime_{\m\n}(X^\prime)]=S[X^\m,g_{\m\n}(X)]$, where
$X^{\prime\,\m}$ and $X^\m$ are related by a target-space general
coordinate transformation which also governs the relation between
$g^\prime_{\m\n}$ and $g_{\m\n}$. Of course, when
considering a background that is invariant under (a subset of 
the) general coordinate transformations (so that $g=g^\prime$), then the
action will be invariant under the corresponding change of the
coordinates. This is the situation that we will address in the next
section, where we take a specific background metric with certain
isometries. In that context the relevant target space for 
\eqn{particle} is an anti-de Sitter ($AdS_d$) space, which has
isometries that constitute the group SO$(d-1,2)$, where 
$d$ is the spacetime dimension. Then \eqn{particle}
describes a one-dimensional field theory with an SO$(d-1,2)$
invariance group. In the particular case of $d=2$ this invariance can be
re-interpreted as a conformal invariance for a supersymmetric quantum
mechanical system.\footnote{%
  This situation arises generically  for any $p$-brane moving in a
  target space that is locally the product of $AdS_{p+2}$ and some
  compact space. The conformal interpretation was emphasized in
  ref. 23. }

Using the previous results one may now write down the complete
action of the supermembrane coupled to background fields up to order
$\theta^2$. Direct substitution leads to the following result for
the supervielbein pull-back,
\bea
\Pi_i^r&=&\partial_iX^\mu\, \Bigl (e_\m{}^r + 2\,
\bar\theta\,\G^r \psi_\m 
-\ft14 \bar\theta\,\G^{rst}\theta\,\hat\omega_{\m\,\!st} + \bar
\theta\,\G^r T_\m{}^{\!\n\rho\s\lambda}\theta \,\hat 
F_{\n\rho\s\lambda} \Big) \nonumber \\
&&
+ \bar\theta\Gamma^r\partial_i\theta + {\cal
O}(\theta^3)\,,\nonumber\\
\Pi_i^a&=& \pa_iX^\m \Big(\psi_\m{}^{\!a} - \ft14\,
\hat\o_\m^{rs}\,(\G_{rs} \theta)^a + 
(T_\m{}^{\!\n\rho\s\lambda}\theta)^a\, \hat
F_{\n\rho\s\lambda}\Big)\nonumber \\
&&+ \pa_i\theta^a  +  {\cal O}(\theta^2)\,.
\label{vielbeinpb}
\eea
Consequently the induced metric is known up to terms of order
$\theta^3$. Furthermore the pull-back of the tensor field equals 
\bea
\lefteqn{-\ft{1}{6}\varepsilon^{ijk}\,
\Pi_i^A\,\Pi_j^B\,\Pi_k^C\, B_{CBA} =  
 -\ft16\vep^{ijk} \pa_iZ^M\,\pa_jZ^N\,\pa_kZ^P\, B_{PNM} =} \nn\\
&& \ft{1}{6}\, dX^{\mu\nu\rho}\, \Big[ 
C_{\m\n\rho} -6\, \bar \theta
\G_{\m\n}\psi_{\rho}
+\ft34\bar\theta\,\G_{rs}\G_{\m\n}\theta\,\hat\omega_{\rho}{}^{\!rs}
\nn\\
&&\hspace{17mm} - 3 \,\bar\theta\,\G_{\m\n}
T_{\rho}{}^{\!\s\lambda\kappa\tau}\theta \,\hat
F_{\s\lambda\kappa\tau} - 12\,
\bar\theta\,\G_{\s\m} \psi_{\n}\;\bar\theta \,\G^\s
\psi_{\rho}\Big]\nonumber\\ 
&&- \varepsilon^{ijk} \,
\bar\theta\,\Gamma_{\mu\nu}\partial_k\theta \Big
[{\textstyle{1\over 2}} 
\,\partial_i X^\mu (\partial_j X^\nu +
\bar\theta\,\Gamma^\nu\partial_j \theta) + {\textstyle{1\over 6}}
\,\bar\theta\,\Gamma^\mu\partial_i\theta\;
\bar\theta\,\Gamma^\nu\partial_j\theta \Big] \nn\\
&& +\ft13 \varepsilon^{ijk} \pa_i X^\m \,\pa_jX^\n \Big[
4\,\bar\theta \,\G_{\rho\m}\pa_k\theta \;
\bar\theta\,\G^\rho\psi_\n - 2 \,\bar \theta\,
\G^\rho\pa_k\theta\; \bar\theta\,\G_{\rho\m}\psi_\n\Big] 
+ {\cal O}(\theta^3)\,, \quad
\label{WZWpb}
\eea
where we have introduced the abbreviation $dX^{\mu\nu\rho}=
\varepsilon^{ijk}\,  
\partial_i X^\mu\,\partial_j X^\nu\, \partial_k X^\rho$ for the
world-volume form. Observe that we included also the 
terms of higher-order $\theta$-terms that were determined in
previous sections and listed in \eqn{stensor}. The first formula
of \eqn{vielbeinpb} and \eqn{WZWpb} now determine the
supermembrane action \eqn{supermem} up to order $\theta^3$. It has
been shown that the resulting expressions are consistent with
supersymmetry and $\kappa$-invariance  \cite{backgr}. 

\section{Near-horizon geometries}
In the previous section we discussed the determination of superspace
quantities,  i.e.  the superspace vielbein and the
tensor gauge field, in terms of the fields of 11-dimensional on-shell
supergravity. The corresponding expressions are obtained by iteration
order-by-order in $\theta$ coordinates, but except for the leading
terms it is hard to proceed with this program. Nevertheless these
results enable one
to write down the 11-dimensional supermembrane action coupled to  
a nontrivial supergravity component-field background to second
order in $\theta$, so that one can start a study of the supermembrane
degrees of freedom in the corresponding  background geometries. 

However, in specific backgrounds with a certain amount of symmetry, it
is possible to obtain results to all orders in $\theta$. 
Interesting candidates for such backgrounds are the
membrane  \cite{sugramem} and the five-brane solution  \cite{sugrafive}
of 11-dimensional supergravity,
as well as solutions corresponding to the product of anti-de Sitter 
spacetimes with compact manifolds. 
Coupling to the latter solutions, which appear near the
horizon of black D-branes  \cite{HorStrom}, seem especially
appealing in view of the recent results on a connection between
large-$N$ superconformal field theories and supergravity on a 
product of  an $AdS$ space with a compact manifold  \cite{maldacena}. The
target-space geometry induced by the $p$-branes 
interpolates between $AdS_{p+2}\times B$ near the horizon, where $B$
denotes some compact manifold (usually a sphere), and flat
$(p+1)$-dimensional Minkowski space times a cone with base $B$.

This program has been carried out recently  for the 
type-IIB superstring and the D3-brane
in a IIB-supergravity background of this type  \cite{MT,KRR,MT2}. In the
context of 11-dimensional  
supergravity the $AdS_4\times S^7$ and $AdS_7\times S^4$ 
backgrounds stand out as they leave all 32 supersymmetries 
invariant  \cite{sevenS,fourS}. These backgrounds  
are associated with the near-horizon geometries corresponding to 
two- and five-brane configurations and thus to possible 
conformal field theories in 3 and 6 spacetime dimensions with 16 
supersymmetries, whose exact nature is not yet completely known. 
In this section we consider the
supermembrane in these two backgrounds  \cite{DWPPS}.  As the
corresponding spaces are local products of 
homogeneous spaces, their geometric information can be extracted 
from appropriate coset representatives leading to standard 
invariant one-forms corresponding to the vielbeine and 
spin-connections. The approach of ref. 32 
differs from that of ref. 33; 
in the latter one
constructs the geometric information    
exploiting simultaneously the $\kappa$-symmetry of the supermembrane 
action, while in ref. 32 
the geometric information is
determined independently from the supermembrane action. The results
for the geometry coincide with those of ref. 34.

As is well known, the compactifications of the theory to
$AdS_4\times S^7$ and  $AdS_7\times S^4$ are
induced by the antisymmetric 4-rank field strength of
M-theory. 
These two compactifications are thus governed by the 
Freund-Rubin field $f$, defined by (in Pauli-K\"all\'en 
convention, so that we  
can leave the precise signature of the spacetime open),  
\be
F_{\mu\nu\rho\sigma}=6 f \,e\, \varepsilon_{\mu\nu\rho\sigma}\,, 
\label{freundrubin}
\ee
with $e$ the vierbein determinant. 
When $f$ is purely imaginary we are dealing with an $AdS_4\times 
S^7$ background while for real $f$ we have an $AdS_7\times S^4$ 
background. The nonvanishing curvature components corresponding to
the 4- and 7-dimensional subspaces are equal to 
\bea
R_{\mu\nu\rho\sigma}&=&- 4 f^2 ( g_{\mu\rho}\,g_{\nu\sigma}
-g_{\mu\sigma}\,g_{\nu\rho})\,,
\nn\\
R_{\mu'\nu'\rho'\sigma'}&=& f^2 ( g_{\mu'\rho'}\,g_{\nu'\sigma'}
-g_{\mu'\sigma'}\,g_{\nu'\rho'}) \,. \label{curvatures}
\label{sugrasol}
\eea
Here $\m,\n,\rho,\s$ and $\m^\prime,\n^\prime,
\rho^\prime ,\s^\prime$ are 4- and 7-dimensional world indices, 
respectively. We also use $m_{4,7}$ for the inverse radii of the 
two subspaces, defined 
by $\vert f\vert^2={m_7}^2=\ft14{m_4}^2$.
The Killing-spinor equations associated with the 32 
supersymmetries in this background take the form
\bea
\Big(D_\m - f\gamma_\m \gamma_5\otimes {\bf 1} \Big) \epsilon = 
\Big(D_{\m^\prime} +\ft12 f{\bf 1}\otimes\gamma_{\m^\prime}^\prime \Big) 
\epsilon = 0  \,, \label{killing-spinors}
\eea
where we make use of the familiar decomposition of the 
(hermitean) gamma matrices $\gamma_\m$ and
$\gamma^\prime_{\m^\prime}$, appropriate to the 
product space of a 4- and a 7-dimensional subspace.
Here $D_\m$ and $D_{\m^\prime}$ denote the covariant derivatives 
containing the spin-connection fields corresponding to SO(3,1) or 
SO(4) and SO(7) or SO(6,1), respectively. 

The algebra of isometries of the $AdS_4\times S^7$ and 
$AdS_7\times S^4$ backgrounds is given by $osp(8|4;{\bf R})$ and 
$osp(6,2|4)$. Their bosonic subalgebra consists of 
$so(8)\oplus sp(4)\simeq so(8)\oplus so(3,2)$ and $so(6,
2)\oplus usp(4)\simeq so(6,2)\oplus so(5)$, respectively. 
The spinors transform in the $(8,4)$ of this 
algebra. 

One may decompose the generators of $osp(8|4)$ or $osp(6,2|4)$ in 
terms of irreducible representations of the bosonic $so(7)\oplus 
so(3,1)$ and $so(6,1)\oplus so(4)$ subalgebras. In that way one
obtains the bosonic (even)  generators $P_r$, $M_{rs}$, which
generate $so(3,2)$ or $so(5)$,  and $P_{r'}$, $M_{r's'}$, which
generate $so(8)$  
or $so(6,2)$. All the bosonic generators are taken antihermitean 
(in the Pauli-K\"all\'en sense). The fermionic 
(odd) generators $Q_{a a'}$ are Majorana spinors, where we denote the
spinorial tangent-space indices by $a,b,\ldots$ and
$a^\prime,b^\prime,\ldots$ for 4- or 7-dimensional indices. 
The commutation relations between even generators are
\be
\begin{array}{rcl}
{[}P_r,P_s{]}&\!=\!&-4f^2 M_{rs}\,,\\[2mm]
{[}P_r,M_{st}{]}&\!=\!&\delta_{rs}\,P_t-\delta_{rt}\,P_s\,,
\\[2mm]
{[}M_{rs},M_{tu}{]}&\!=\!&\delta_{ru}\,M_{st}+\delta_{st}\,
M_{ru}\\[1mm]
&&-\delta_{rt}\,M_{su}-\delta_{su}\,M_{rt}\,,
\end{array}
\begin{array}{rcl}
{[}P_{r'},P_{s'}{]}&\!=\!&f^2 M_{r's'}\,,\\[2mm]
{[}P_{r'},M_{s't'}{]} &\!=\!& \delta_{r's'}\,P_{t'}-\delta_{r't'}\,
P_{s'}\,,\\[2mm]
{[}M_{r's'},M_{t'u'}{]}&\!=\!&\delta_{r'u'}\,M_{s't'}+\delta_{s't'}\,
M_{r'u'}\\[1mm] 
&& -\delta_{r't'}\,M_{s'u'}-\delta_{s'u'}\,M_{r't'}\,.
\end{array} \label{bosonic-comm}
\ee
The odd-even commutators are given by
\be
\begin{array}{rcl}
{[}P_r,Q_{a a'}{]}&\!=\!&- {f} (\gamma_r\gamma_5)_a {}^b \,Q_{b 
a'}\,,\\[2mm]
{[}M_{rs},Q_{a a'}{]}&\!=\!& -\ft12(\gamma_{rs})_a {}^b \,Q_{ba 
'}\,,
\end{array}
\quad
\begin{array}{rcl}
{[}P_{r'},Q_{a a'}{]}&\!=\!& -\ft12 {f}(\gamma'{}_{\!r'})_{a'} {}^{b'} 
\,Q_{a b'}\,,\\[2mm]
{[}M_{r's'},Q_{a a'}{]}&\!=\!& -\ft12 (\gamma'{}_{\!r's'})_{a'} {}^{b'} 
\,Q_{a b'}\,.
\end{array}
\ee
Finally, we have the odd-odd anti-commutators, 
\begin{eqnarray}
\{Q_{a a '},Q_{bb'}\}&=&-(\gamma_5 C)_{a b}
\left(2(\gamma'{}_{\!r'}C')_{a 'b'}\,P^{r'}
- f(\gamma'{}_{\!r's'}C')_{a 'b'}\,M^{r's'}\right) \nonumber\\
&&-C'{}_{\!a' b'}\Bigl(2(\gamma{}_{r}C)_{a b}\,P^{r}
+ 2  f (\gamma{}_{rs}\gamma_5 C)_{a b}\,M^{rs}\Bigr ).
\end{eqnarray}
All other (anti)commutators vanish. The normalizations of the
above algebra were determined by comparison with the 
supersymmetry algebra in the conventions of   \cite{backgr} 
in the appropriate backgrounds. 

However, one can return to 11-dimensional notation and drop the
distinction between 4- and 7-dimensional indices so that the equations
obtain a more compact form. In that case the above (anti)commutation
relations that involve the supercharges can be concisely written as,
\begin{eqnarray}
{[}P_{  r},\bar Q{]}&\!=\!& \bar Q \, T_{  r}{}^{\!  s 
  t  u  v} F_{  s  t  u  v} \,,\qquad
{[}M_{  r  s},\bar Q{]}= \ft12 \bar Q\,\Gamma_{  r  s} \,, 
\nonumber\\
\{Q,\bar Q\}&\!=\!&-2 \Gamma_{  r}\,P^{  r}
+ \ft1{144} \Big[ \Gamma^{  r  s  t  u  v  w} 
F_{  t  u  v  w} 
+ 24 \,\Gamma_{  t  u}   F^{  r  s  t  u} \Big] 
M_{  r  s} \,, \label{11-comm}
\end{eqnarray}
where the tensor $T$ equals 
\be
T_r{}^{\!stuv} = \ft1{288} \Big (\Gamma_r{}^{\!stuv} - 8\,
\d_r^{[s}\,\Gamma^{tuv]} \Big)\,. 
\ee
Note, however, that the above 
formulae are only applicable in the background  where the field
strength takes the form given in \eqn{freundrubin}. In what
follows, we will only make use of \eqn{11-comm}. 

\section{Coset-space representatives of $AdS_4\times S^7$ 
and $AdS_7\times S^4$} 
\noindent
Both backgrounds that we consider correspond to homogenous spaces 
and can thus be formulated as coset spaces. In the case at hand 
these (reductive) coset spaces $G/H$ are $OSp(8|4;{\bf 
R})/SO(7)\times SO(3,1)$ and $OSp(6,2|4)/SO(6,1)\times SO(4)$. 
To each element of the coset $G/H$ one associates an element of 
$G$, which we denote by $L(Z)$. Here $Z^A$ stands for the 
coset-space coordinates $x^{  r}$, $\theta^{  a}$ (or, 
alternatively, $x^r$, $y^{r'}$ and $\theta^{a a '}$). The 
coset representative $L$ transforms from the left under constant 
$G$-transformations corresponding to the isometry group of the 
coset space and from the right under local $H$-transformations: 
$L\to L^\prime = g\,L\,h^{-1}$. 

The vielbein and the torsion-free $H$-connection one-forms, $E$ 
and $\Omega$, are defined through\footnote{%
  A one-form $V$ stands for $V\equiv {\rm d}Z^AV_A$ and an exterior 
  derivative acts according to ${\rm d}V\equiv -{\rm d}Z^B\wedge 
  {\rm d}Z^A \, \partial_AV_B$. Fermionic derivatives are thus 
  always left-derivatives.} 
\be
{\rm d} L + L\,\Omega = L\, E\,,\label{defv}
\ee
where 
\begin{eqnarray}
E= E^{  r}P_{  r} +\bar E Q \,,\qquad 
\Omega= \ft12 \Omega^{  r  s}M_{  r   s}.
\end{eqnarray}
The integrability of \eqn{defv} leads to the Maurer-Cartan 
equations, 
\begin{eqnarray}
\label{maurercartan}
&&{\rm d}\Omega - \Omega \wedge \Omega - \ft12  E^{  r}\wedge 
E^{  s} \, [P_{  r},P_{  s}] \nonumber\\
&&\hspace{1cm} - \ft1{288} \bar E\Big[ \Gamma^{  r  s  t  u  v  w} 
F_{  t  u  v  w} 
+ 24 \,\Gamma_{  t  u}   F^{  r  s  t  u} \Big] 
E\,M_{  r  s} =0  \,, \nn\\
&&{\rm d} E^{  r} -\Omega^{  r  s}\wedge E_{  s} - 
\bar E\,\Gamma^{  r} \wedge E=0 \, ,\nn \\
&&{\rm d} E +  E^{  r}\wedge T_{  r}{}^{  t  u  
v  w}E \,F_{  t   u   v   w} - 
\ft14 \Omega^{  r  s} \wedge \Gamma_{  r  s} E=0 \, ,
\end{eqnarray}
where we suppressed the spinor indices on the anticommuting 
component $E^{  a}$. 
The first equation in a fermion-free background reproduces 
\eqn{curvatures} upon using the commutation relations 
\eqn{bosonic-comm}.   

Now the question is how to determine the vielbeine and
connections to all orders in $\theta$ for the spaces of interest.  
First, observe that the choice of the coset representative amounts to
a gauge choice  
that fixes the parametrization of the coset space. We will 
not insist on an explicit parametrization of the bosonic part of 
the space. It turns out to be advantageous to factorize $L(Z)$ 
into a group element of the bosonic part of $G$ corresponding 
to the bosonic coset space, whose parametrization we leave 
unspecified, and a fermion factor. Hence one may write 
\be
L(Z) =  \ell(x)\, \hat L(\theta)\,, \quad \mbox{ with } \hat 
L(\theta) = \exp [\,\bar \theta Q\,]\,. 
\ee
There exists a convenient trick  \cite{MT,KRR,ssp} according to
which one first rescales the odd  
coordinates according to $\theta\rightarrow t\, \theta$, where $t$ is 
an auxiliary parameter that we will put to unity at the end. 
Taking the derivative with respect to $t$ of \eqn{defv} then 
leads to a first-order differential equation for $E$ and $\Omega$ (in 
11-dimensional notation),
\begin{eqnarray}
\dot E - \dot \Omega &=& {\rm d}\bar \theta \,Q  + (E-\Omega) \,
\bar\theta Q - \bar\theta Q \,(E-\Omega) 
\end{eqnarray}
After expanding $E$ and $\Omega$ on the right-hand side in 
terms of the generators and using the (anti)commutation relations 
\eqn{11-comm} we find a set of coupled first-order linear differential
equations,  
\begin{eqnarray}
\dot E^{  a} &=& \Bigl({\rm d}\theta +  E^{  r} \,
T_{  r}{}^{\!  s   t  u  v} \theta \,
F_{  s  t  u  v}  -  \ft14 \Omega^{  r  s}\,
\Gamma_{  r  s}\theta \Bigr)^{  a}\,,\nonumber \\ 
\dot E^{  r} &=& 2\, \bar \theta \,\Gamma^{  r} E\,,\nonumber 
\\
\dot \Omega^{  r  s} &=&  \ft1{72} \bar \theta \Big[ 
\Gamma^{  r  s  t  u  v  w}  
F_{  t  u  v  w} 
+ 24 \,\Gamma_{  t  u}   F^{  r  s  t  u} \Big] 
E\,. \label{diffeq}
\end{eqnarray}
These equations can be solved straightforwardly  \cite{KRR} and one finds
\bea
E(x,\theta)&=&\sum_{n=0}^{16}\, \frac{1}{(2n+1)!}\, {\cal M}^{2n} 
\, D\theta\,,  \nn \\ 
E^{  r}(x,\theta)&=&{\rm d}x^{  \mu}\, e_{  \mu} 
{}^{\!  r}  + 2 \sum_{n=0}^{15}\,\frac{1}{(2n+2)!}\, 
\bar\theta\,\Gamma^{  r}\, {\cal M}^{2n}\, D\theta
\\[2.5 mm]
\Omega^{  r  s}(x,\theta)&=&{\rm d}x^{  \mu}\,  
\omega_{  \mu}{}^{\!  r  s}  \nn \\
&& +\ft1{72}  
\sum_{n=0}^{15}\frac{1}{(2n+2)!}\, \bar\theta \,[ 
\Gamma^{  r  s  t  u  v  w}  
F_{  t  u  v  w} 
+ 24 \,\Gamma_{  t  u}   F^{  r  s  t  u} ] 
\, {\cal M}^{2n}\, D\theta    \,, 
\nn
\eea
where the matrix ${\cal M}^2$ equals,
\bea
({\cal M}^2)^{  a}{}_{\!  b}  &=& 2\, (T_{  r}{}^{\!  s 
  t  u  v}\, \theta )^{  a}\, 
F_{s  t  u  v} \, (\bar \theta \,\Gamma^{  
r})_{b} \nn\\
&& - \ft1{288} (\Gamma_{  r  s}\, \theta)^{  
a}\, (\bar\theta\,[\Gamma^{  r  s  t  u  v  w}  
F_{t  u  v  w} 
+ 24 \,\Gamma_{t  u}   F^{  r  s  t  
u}])_{  b}\,.
\eea
and
\be
D\theta^{  a} \equiv \dot E^{  a}\Big\vert_{t=0} =  \Big({\rm d}\theta +  e^{  r} \,
T_{  r}{}^{\!  s   t  u  v} \theta  \, 
F_{  s  t  u  v}  -  \ft14 \omega^{  r  s}\,
\Gamma_{  r  s}\theta\Big)^{  a}\,. 
\ee
It is straightforward to write down the lowest-order terms in these
expansions, 
\begin{eqnarray}
\label{vielbeinexp}
E^{  r} &=& e^{  r} + \bar\theta\Gamma^{  r} {\rm 
d}\theta +  
\bar\theta \Gamma^{  r} ( e^{  m}\,T_{  m}{}^{  s  
t  u  v}  
  F_{  s  t  u  v} - \ft14 \omega^{  s  t} \,
\Gamma_{  s  t})\theta 
+ {\cal O}(\theta^4)\, ,\nn \\
E &=& {\rm d}\theta + 
( e^{  r}\,T_{  r}{}^{  s  t  u  v}
F_{  s  t  u   v} - \ft14 \omega^{  r  s}\, 
\Gamma_{  r  s}) \theta + {\cal O}(\theta^3)   \,,\nn \\
\Omega^{  r  s} &=& \omega^{  r   s}  +\ft1{144}  
 \bar\theta \,[ 
\Gamma^{  r  s  t  u  v  w}  
F_{  t  u  v  w} + 24 \,\Gamma_{  t  u}  
F^{  r  s  t  u} ] \, {\rm d}\theta  
+ {\cal O}(\theta^4) \, , 
\end{eqnarray}
which agree completely with those given in section~8 (and, for the
spin-connection field, in ref. 17).

This information can now be substituted into the first part of the
supermembrane action \eqn{supermem}. 
By similar techniques one can also determine the Wess-Zumino-Witten
part of the action by  first 
considering the most general ansatz for a four-form invariant 
under tangent-space transformations. Using the lowest-order 
expansions of the vielbeine 
(\ref{vielbeinexp}) and comparing with ref. 18 
shows that only two terms can be present. Their relative coefficient
is fixed  by requiring that the four-form is closed, something that
can be verified by making use of the Maurer-Cartan equations 
(\ref{maurercartan}). The result takes the form 
\begin{equation}
\label{wzwfourform}
F_{(4)} = \frac{1}{4!}\Big[ 
E^{  r}\wedge E^{  s} \wedge E^{  t} \wedge E^{  u} 
F_{  r  s  t  u} 
- 12 \, \bar E \wedge \Gamma_{  r  s} E \wedge E^{  r} 
\wedge E^{  s} \Big]\, .
\end{equation}
To establish this result we also needed the well-known 
quartic-spinor identity in 11 dimensions. The overall factor in 
\eqn{wzwfourform} is fixed by comparing to the normalization of 
the results given in ref. 18.

Because $F_{(4)}$ is closed, it can be written locally  as  
$F_{(4)}= {\rm d} B$. 
The general solution for $B$ can be found by again exploiting the 
one-forms with rescaled $\theta$ coordinates according to $\theta\to t\,
\theta$ and deriving a differential equation. Using the lowest order
result for $B$ this equation can be solved and yields  
\be
B= \ft16\, e^{  r}\wedge e^{  s}\wedge e^{  t} \,C_{  
r   s   t} 
-\int_0^1{\rm d}t\;  \bar\theta \,
\Gamma_{  r  s} E \wedge E^{  r}\wedge E^{  s}\, ,
\ee
where the vielbein components contain the rescaled $\theta$'s. 
This answer immediately reproduces the flat-space result upon
substitution of $F_{r stu}= \omega^{rs}=0$. 

In order to obtain the supermembrane action one substitutes the above
expressions in the action \eqn{supermem}. The resulting action is then
invariant under local fermionic $\kappa$-transformations as
well as under the superspace isometries  
corresponding to $osp(8\vert 4)$ or $osp(6,2\vert 4)$.

We have already emphasized that the choice of the coset representative 
amounts to adopting a certain gauge choice in superspace. The choice
that was made in ref. 32 
connects directly to the generic 
11-dimensional superspace results, written in a Wess-Zumino-type 
gauge, in which there is no distinction between spinorial 
world and tangent-space indices. In specific backgrounds, such as 
the ones discussed here, gauge choices are possible which 
allow further simplifications. For this we refer to refs. 28 and 36.

The results of this section provide a convincing check of the
low-order $\theta$ results obtained by gauge completion for general
backgrounds  \cite{backgr,CF}. 
A great amount of clarity was gained by expressing our results 
in 11-dimensional language, so that both
the $AdS_4\times S^7$ and the $AdS_7\times S^4$ solution could be
covered in one go. Note that in both these  backgrounds the 
gravitino vanishes. 

We have no reasons to expect that the 
11-dimensional form of our results will coincide with the 
expressions for a generic 11-dimensional superspace (with the 
gravitino set to zero) at arbitrary orders in $\theta$.

From a more technical viewpoint it is gratifying that explicit
constructions of supermembranes in certain nontrivial backgrounds are
now possible. The complete 
M-theory two-brane action in $AdS_4\times S^7$ and $AdS_7\times S^4$
to all orders in $\theta$ represents a further step
in the program of finding the complete anti-de-Sitter background actions 
for the superstring  \cite{MT,KRR} and the M2-, D3-  \cite{MT2} and
M5-branes initiated for the bosonic part in ref. 23. Furthermore,
by studying the interaction between a test membrane in the background
of an M2- or an M5-brane, one may hope to learn more about the
interactions between branes. Some of these issues have already been
considered recently  \cite{interactions}.  

\section*{Acknowledgments}
The results described in this talk were obtained in collaboration with
Kasper Peeters, Jan Plefka and Alexander Sevrin.



\begin{thebibliography}{99}  
%
\bibitem{BST} E. Bergshoeff, E. Sezgin and P.K. Townsend, 
Phys. Lett. {\bf 189B} (1987) 75;   Ann. Phys. {\bf 185} (1988) 330. 
%
\bibitem{CJS} E. Cremmer, B. Julia and J. Scherk, Phys. Lett. 
{\bf 76B} (1978) 409. 
%
\bibitem{GS84} M.B. Green and J.H. Schwarz, \PLB{136}{1984}{367};
\NPB{243}{1984}{285}.
%
\bibitem{DWLN}
 B. de Wit, M. L\"uscher and H. Nicolai, {Nucl. Phys.} {\bf 
B320} (1989) 135.
%
\bibitem{DWHN} B. de Wit, J. Hoppe, H. Nicolai, Nucl. Phys. 
{\bf B305} [FS23] (1988) 545.
%
\bibitem{DWN} B.\ de Wit and H.\ Nicolai, in proc.\ Trieste
Conference on
Supermembranes and Physics in $2+1$ dimensions, p.\ 196, eds.\ M.J.
Duff,
C.N.\ Pope and E.\ Sezgin (World Scient., 1990);\\
B.~de~Wit,  {Nucl.\ Phys. B (Proc.\
Suppl.)} {\bf 56B} (1997) 76, {\tt  hep-th/9701169}.
%
\bibitem{HT}
C.M. Hull and P.K. Townsend, {Nucl. Phys} {\bf B438} (1995)
{109}, {\tt hep-th/9410167}.
%
\bibitem{Townsend} P.K.\ Townsend, {Phys.\ Lett.}\ {\bf B350} 
(1995) 184, {\tt hep-th/9501068}, {Phys.\ Lett.}\ {\bf B373} 
(1996) 68, {\tt hep-th/9512062}.
%
\bibitem{witten3} E. Witten, {Nucl.\ Phys.} {\bf B443} (1995)
85, {\tt hep-th/9503124}.
%
\bibitem{horvw} P.\ Ho\u{r}ava and E.\ Witten,
{Nucl.\ Phys.} {\bf B460} (1996) {506}, {\tt hep-th/9510209}, {\bf
B475} (1996) {94}, {\tt hep-th/9603142}.
%
\bibitem{CH} M. Claudson and M.B. Halpern, {Nucl.\ Phys.} {\bf
B250} (1985) 689;\\
R. Flume, {Ann.\ Phys.} {\bf 164} (1985) 189;\\
M. Baake, P. Reinicke, and V. Rittenberg, {J. Math.\ Phys.}
{\bf 26.} (1985) 1070. 
%
\bibitem{boundst} E.\ Witten, Nucl.\ Phys.\ {\bf B460} 
(1996) 335, {\tt hep-th/9510135}.  
%
\bibitem{BFSS} T.\ Banks, W.\ Fischler, S.H.\ Shenker and L.\ Susskind, 
\PRD{55}{1997}{5112}, {\tt hep-th/9610043}.
%
\bibitem{mboundstates}
P. Yi, \NPB{505}{1997}{307}, {\tt hep-th/9704098};\\
S. Sethi and M. Stern, 
Commun. Math. Phys. {\bf 194} (1998) 675, 
{\tt hep-th/9705046};\\
M. Porrati and A. Rosenberg, 
Nucl. Phys. {\bf B515} (1998) 184, {\tt hep-th/9708119}.
%
\bibitem{oldBG} 
M.R. Douglas, H. Ooguri and S.H. Shenker, {Phys.\ Lett.} {\bf
B402} (1997) 36, {\tt hep-th/9702203};\\ 
M.R. Douglas, {\it D-branes in curved space}, {\tt hep-th/9703056};\\
M.R. Douglas, A. Kato and H. Ooguri, {\it D-brane actions on
K\"ahler manifolds},  {\tt hep-th/9708012};\\
M.R. Douglas and H. Ooguri, 
Phys. Lett. {\bf B425} (1998) 71, {\tt hep-th/9710178}. 
%
\bibitem{maldacena} J. Maldacena, 
Adv. Theor. Math. Phys. {\bf 2} (1998) 231, {\tt hep-th/9711200};\\
S.S. Gubser, I.R. Klebanov and A.M. Polyakov, \PLB{428}{1998}{105}, {\tt 
hept-th/9802109};\\
E. Witten, 
Adv. Theor. Math. Phys. {\bf 2} (1998) 253, {\tt hep-th/9802150}.
%
\bibitem{CF} E. Cremmer and S. Ferrara, {Phys.\ Lett.} {\bf
91B} (1980) 61. 
%
\bibitem{backgr} B. de Wit, K. Peeters and J.C. Plefka, Nucl. 
Phys. {\bf B532} (1998) 99, {\tt hep-th/9803209}.
%
\bibitem{GS} M. Goroff and J.H. Schwarz, {Phys.\ Lett.} {\bf 
127B} (1983) 61. 
%
\bibitem{DWPP3} B. de Wit, K. Peeters and J.C. Plefka,
in proc. of the Val\`encia 97 workshop {\it Beyond the standard
model; from theory to experiment}, eds. I. Antoniadis, L.E. Ibanez and
J.W.F. Valle, World Scient. 1998,  {\tt hep-th/9712082}.
%
\bibitem{newBG} A. Connes, M.R. Douglas and A. Schwarz, 
JHEP {\bf 02} (1998) 003, {\tt hep-th/9711162};\\ 
M.R. Douglas and C. Hull, 
JHEP {\bf 02} (1998) 008, {\tt hep-th/9711165};\\
N.A. Obers, B. Pioline and E. Rabinovici, 
Nucl. Phys. {\bf B525} (1998) 163, {\tt hep-th/9712084}.
%
\bibitem{interactions}
V. Balasubramanian, D. Kastor, J. Traschen and K.Z. Win, {\it The spin  
of the M2-brane and spin-spin interactions via probe techniques}, {\tt
hep-th/9811037};\\
W. Taylor and M. Van Raamsdonk, {\it Supergravity currents and
linearized interactions for Matrix Theory configurations with
fermionic backgrounds}, {\tt hep-th/9812239};\\
T. Sato, {\it Superalgebras in many types of M-brane backgrounds and
various supersymmetric brane configurations}, {\tt hep-th/9812014};\\
S. Hyun, Y. Kiem and H. Shin, {\it Non-perturbative  membrane
spin-orbit couplings in M/IIA theory}, {\tt hep-th/9901105}.
%
\bibitem{KvPTK} P. Claus, R. Kallosh, J. Kumar, P. Townsend, A. Van Proeyen,
\JHEP{06}{1998}{004}, {\tt hep-th/9801206}.
%
\bibitem{sugramem}
M.J. Duff and K.S. Stelle, \PLB{253}{1991}{113}.
%
\bibitem{sugrafive}
R. G\"uven, \PLB{276}{1992}{49}.
%
\bibitem{HorStrom} G. Horowitz and A. Strominger, {Nucl. 
Phys.} {\bf B360} (1991) 197. 
%
\bibitem{MT} R.R. Metsaev and A.A. Tseytlin, 
Nucl. Phys. {\bf B533} (1998) 109, {\tt hep-th/9805028}, 
%
\bibitem{KRR} R. Kallosh, J. Rahmfeld and A. Rajaraman, 
JHEP {\bf 09} (1998) 002, {\tt hep-th/9805217}.
%
\bibitem{MT2} R.R. Metsaev and A.A. Tseytlin, 
Phys. Lett. {\bf B436} (1998) 281, {\tt hep-th/9806095}. 
%
\bibitem{sevenS} M.J. Duff and C.N. Pope, {\it in} Supersymmetry 
and Supergravity '82,  eds. S. Ferrara, J.G. Taylor and P. van 
Nieuwenhuizen  (World Scientific, 1983), 183; \\
B. Biran, F. Englert, B. de Wit and H. Nicolai,
Phys. Lett. {\bf 124B} (1983) {45}; (E) 128B (1983) 461. 
%
\bibitem{fourS} K. Pilch, P. van Nieuwenhuizen and P.K. 
Townsend, Nucl. Phys. {\bf B242} (1984) 377. 
%
\bibitem{DWPPS} B. de Wit, K. Peeters, J. Plefka and A. Sevrin,
 Phys. Lett. {\bf B443} (1998) 153, {\tt hep-th/9808052}. 
%
\bibitem{DFFFTT} G. Dall'Agata, D. Fabbri, C. Fraser, P. Fr\'e, 
P. Termonia and M. Trigiante, {\it The $Osp(8\vert4)$ singleton 
action from the supermembrane}, {\tt hep-th/9807115}.
%
\bibitem{Claus} P. Claus, {\it Super M-brane actions in $AdS_4\times
S^7$ and $AdS_7\times S^4$},  {\tt hep-th/9809045}. 
%
\bibitem{ssp} D.G. Boulware and L.S. Brown, Ann. Phys. {\bf 138} 
(1982) 392. 
%
\bibitem{kallosh} R. Kallosh, {\it Superconformal actions in  
Killing gauge}, {\tt hep-th/9807206}.

\end{thebibliography}
\end{document}